\documentclass[showpacs,preprintnumbers,twocolumn]{revtex4}
\usepackage{graphicx, epsfig}
\usepackage{bm}

\begin{document}

\preprint{SU-GP-02/10-3}
\preprint{SU-4252-770}
\preprint{CWRU-P41-02}
\title{A Model for Neutrino Masses and Dark Matter}

\author{Lawrence M. Krauss$^1$, Salah Nasri$^2$ and Mark Trodden$^2$}

\affiliation{$^1$Departments of Physics and Astronomy \\
Case Western Reserve University \\
10900 Euclid Ave. \\
Cleveland, OH 44106-7079, USA \\
\\
$^2$Department of Physics \\
Syracuse University \\
Syracuse, NY 13244-1130, USA}

\date{\today}

\begin{abstract}
We propose a model for neutrino masses that simultaneously results in a new dark matter candidate, the right-handed neutrino. We derive the dark matter abundance in this model, show how the hierarchy of neutrino masses is obtained, and verify that the model is compatible with existing experimental results. The model provides an economical method of unifying two seemingly separate puzzles in contemporary particle physics and cosmology.
\end{abstract}


\maketitle

\section{Introduction}
\label{intro}
The discovery of very small, but non-zero neutrino masses via the observation of oscillations in atmospheric and solar neutrino experiments~\cite{superk,SNO} is one of the most exciting results in particle physics in recent years, and may provide important new information to help guide us in the search for new physics beyond the standard model.   As we describe here, efforts to uncover a mechanism for generating such small mass differences may also shed light on the nature of non-baryonic dark matter.

The simplest way to introduce neutrino masses into the standard model is to postulate a new right-handed fermion, a singlet under the SM gauge group, that mixes with the SM fermions via a Dirac mass term. Unfortunately, in order to obtain neutrino masses below an eV, it is necessary to fine-tune the corresponding Yukawa coupling to one part in $10^{11}$, which is unacceptable. Thus, more elaborate models are needed.

Currently the most popular way to generate small neutrino masses
is the \textit{seesaw} mechanism~\cite{seesaw}. In this model one introduces a right handed
neutrino $N_R$, and allows both Majorana and Dirac mass terms. Since
the right handed neutrino is a singlet under the SM gauge group $SU(2)_L \times U(1)_Y$,
its mass $M_R$ can be much larger than the electroweak scale without contributing large quantum corrections to the other couplings in the theory. One may integrate out such a heavy field leaving a Majorana neutrino with mass $m_\nu \sim \frac{m_D^2}{M_R}$,
where $m_D$ is the Dirac neutrino mass. The seesaw mechanism arises naturally
in many extensions of the standard model gauge group, for example in left-right (L-R) models with group $SU(3)_c\times SU(2)_L\times SU(2)_R \times U(1)_{B-L}$, the Pati-Salam model with group $SU(4)_c \times SU(2)_L\times SU(2)_R$ or in $SO(10)$ grand unified theories (GUTs).

Another possible way of achieving small neutrino masses is to extend the Higgs sector of the theory by adding a very heavy triplet scalar field $\Delta$, with mass $M_{\Delta}$, which couples to the SM fields via
\begin{equation}
\label{LDelta}
{\cal L}_{\Delta} = f_{\alpha\beta}L^{T}_{\alpha}Ci\tau_2 \Delta L_{\beta} + \mu\Phi^T i\tau_2 \Delta^+ H\Phi+ {\rm h.c.} 
\ ,
\end{equation}
where, $L_{\alpha}$ is the left-handed lepton doublet, $\Phi$ is the standard model Higgs doublet and $C$ is the charge-conjugation matrix. Here the Yukawa couplings $f_{\alpha\beta}$ are antisymmetric in the generation indices $\alpha$ and $\beta$.

Integrating out our heavy triplet $\Delta$ one obtains an effective theory whose Lagrangian density contains the term
\begin{equation}
{\cal L}_{LH} = \frac{1}{\Lambda}LLHH \ ,
\end{equation}
where the scale $\Lambda \sim M_{\Delta}^{-2}$. Due to this term, after electroweak symmetry
breaking the neutrino gets a mass $m_\nu \sim \frac{V_{EW}^2}{M_{\Delta}}$ where $V_{EW}$ is the vacuum expectation value (VEV) of the standard model higgs field.

Clearly, in both of these sets of models for generating neutrino masses, it is crucial that the masses of the right-handed neutrino and the scalar triplet respectively are much higher than the electroweak scale. As a consequence, these models are difficult to test in collider experiments. In addition, the existence of a large hierarchy between the electroweak scale and the new physics scale generates its own set of naturalness problems.

Here we reconsider the idea that SM neutrinos are massless at tree level, but acquire a nonzero mass at the one-loop level. A particularly simple example of this idea is provided by the
Zee model~\cite{zee} in which one augments the SM higgs doublet $\Phi_1$ with a second higgs doublet $\Phi_2$, and a charged field $S$ which transforms as a singlet under $SU(2)$. The Lagrangian density then contains the following relevant pieces
\begin{equation}
{\cal L}_{\rm Zee} = f_{\alpha\beta}L_{\alpha}^TCi\tau_2L_{\beta}S^{+} 
+ \mu \Phi_1^Ti\tau_2\Phi_2 S^{-} + {\rm h.c.} \ .
\end{equation}
Other extensions of the Zee model where small neutrino masses
arise at loop levels have also been proposed~\cite{Nasri:2001ax} that may have interesting implications for the generation of the baryon asymmetry of the universe~\cite{Nasri:2001nb, Luty:un}.

In this letter we will demonstrate how a variant of the Zee model allows the generation of small phenomenologically acceptable neutrino masses and at the same time produces a new viable dark matter candidate near the electroweak scale.  

\section{The Model}
\begin{figure}[h]
\epsfxsize = 1.0 \hsize \epsfbox{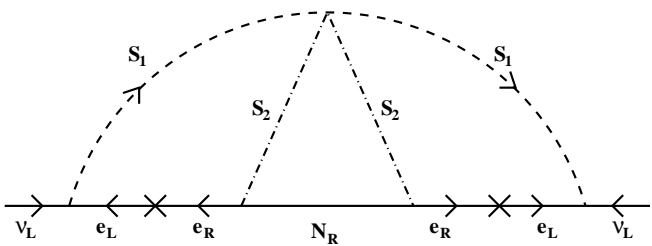}
\caption{\label{figure1}
The three-loop diagram through which a nonzero neutrino mass is generated.}
\end{figure}

We consider a model with sufficient symmetry that neutrino masses appear only at the three loop level. To achieve this we supplement the SM fields with two charged scalar singlet scalars $S_1$ and 
$S_2$, with masses $M_{S_1}$ and $M_{S_2}$ respectively
and one right handed neutrino $N_R$, with mass $M_R$. We break lepton number explicitly by including a Majorana mass term for the right-handed neutrino, and impose a discrete $Z_2$ symmetry under which the SM fields and $S_1$ are singlets but $S_2$ and $N_R$ transform as
\begin{equation}
Z_2: \{S_2, N_R\} \longrightarrow \{-S_2, -N_R\} \ ,
\end{equation}
forbidding Dirac masses for the neutrinos. Given this symmetry, the most general renormalizable terms  that may be added to the SM Lagrangian density
\begin{eqnarray}
{\cal L}_{\rm new} = && f_{\alpha\beta}L_{\alpha}^TCi\tau_2L_{\beta}S_1^{+} +
g_{\alpha}N_RS_2^{+}l_{{\alpha}_R} \nonumber \\
&& + M_RN_R^TCN_R + V(S_1,S_2)
+ {\rm h.c.} \ ,
\end{eqnarray}
in which the potential $V(S_1,S_2)$ contains a $(S_1S_2^{*})^2$ coupling $\lambda_s$.
Let us assume a mild hierarchy of masses $M_R<M_{S_1}<M_{S_2} \sim $ TeV and that the
Yukawa couplings $f_{\alpha\beta}$, $g_{\alpha}$ are of order one. This implies that $N_R$ is
stable if the above discrete symmetry is unbroken. Also, since lepton number is broken, we expect a left handed Majorana neutrino mass at the quantum level, and indeed this arises
from the three loop diagram in figure~{\ref{figure1}}. 

Once again, we integrate out the heavy degrees of freedom $N_R$, $S_1$ and $S_2$ to obtain an effective theory containing the dimension five operator
\begin{equation}
{\cal L}_{LH} = \frac{1}{\Lambda}(L^T_{\alpha}C
\overline{\tau}\Phi_1)(\phi_1^T\overline{\tau}L_{\alpha}) \ .
\end{equation}
Here the scale $\Lambda$ is given in terms of the heaviest singlet
scalar mass $M_{S_2}$ and the Yukawa couplings $\lambda_l = \frac{m_l}{V_{EW}}$,
$f_{\alpha\beta}$, $g_{\alpha}$ and $\lambda_s$ as
\begin{equation}
\Lambda \sim \frac{(4\pi^2)^3}{f^2\lambda_s{\lambda_l}^2 g_{\alpha}^2} M_{S_2} \ ,
\end{equation}
where a subscript $l$ denotes a leptonic quantity.
For $M_{S_2} \sim$ TeV, we obtain $\Lambda > 10^9$ GeV and so the model provides a hierarchy between the electron and neutrino masses and yields neutrino masses at the $0.1$ eV scale without involving mass scales significantly larger than a TeV.

\section{Dark Matter}
As we commented above, in our model the right-handed neutrino is stable, since its mass is smaller than
$M_{S_1}$ and the discrete symmetry $Z_2$ is unbroken. It is natural to wonder whether this stable, neutral, weakly-interacting particle might be a reasonable cold dark matter (CDM) candidate.  To determine this we need to calculate the relic abundance of right-handed neutrinos produced as the universe cools. 

We begin by noting that the right-handed neutrino annihilates into right-handed charged leptons with cross-section given by
\begin{equation}
\langle\sigma v\rangle \sim \frac{g^4}{\pi}\frac{M_R^2}{M_{S_2}^4} \ .
\end{equation}
We also need the standard result that the number density of right-handed neutrinos at the decoupling temperature $T_D$ is
\begin{equation}
n_{N_R+{\bar N}_R} = \frac{2 T_D^3}{(2\pi)^{3/2}}\left(\frac{M_R}{T_D}\right)^{3/2} \exp\left(-\frac{M_R}{T_D}\right) \ .
\end{equation}
Freeze-out occurs when
\begin{equation}
n_{N_R}\sigma \simeq \sqrt{g_*}\frac{T_D^2}{M_{pl}} \ ,
\end{equation}
where $g_*$ is the effective number of massless degrees of freedom of at decoupling. Therefore, using our expressions for $n_{N_R+{\bar N}_R}$ and the
cross-section we solve for $\frac{M_R}{T_D}$ to obtain
\begin{eqnarray}
\frac{M_R}{T_D} =
&-& \ln\left(\frac{1.66(2\pi)^{\frac{3}{2}}M_{S_2}^4}{M_{pl}g^4{\rm GeV}^3}\right) -
\frac{1}{2}\ln\left(\frac{M_R}{T_D}\right) \nonumber \\
&+& 3\ln\left(\frac{M_R}{{\rm GeV}}\right) -
\ln\left(\frac{\sqrt{g_*}}{g_R}\right) \ ,
\end{eqnarray}
where the minus sign in front of the second term arises because the annihilation of $N_R$ is a p-wave process, and where we can ignore the masses of the annihilation products with masses less than $M_R$. For typical values $g^2 \sim 0.1$ and $M_{S_2} \sim$ TeV we obtain
\begin{equation}
\frac{M_R}{T_D} \simeq 20 \ .
\end{equation}

The ratio of the energy density in right-handed neutrinos to the critical density is approximately given by
\begin{equation}
\Omega_{N_R}\simeq \frac{10^{-37}{\rm cm}^2}{\langle\sigma v \rangle} \ .
\end{equation}
Note that if we take typical values $g^2\sim 0.1$, $M_R\sim$ TeV then $ \langle\sigma
v\rangle \sim 10^{-36}{\rm cm}^2$ and so one quite easily finds
\begin{equation}
\Omega_{N_R}\simeq 0.1 - 1 \ ,
\end{equation}
as required for a viable CDM candidate.

Note also that because the right handed neutrino has no direct neutral current couplings to
quarks, it would not be detected in existing WIMP detection experiments sensitive to elastic
scattering off of nuclei.

\section{Current Experimental Constraints}
We next consider whether what current phenomenological constraints on lepton number violation imply for this model.

First note that stringent constraint on the parameters $f_{\alpha\beta}$ can be
derived from the new contribution to the muon decay effective
lagrangian mediated by $S_1$. We find
\begin{equation}
\frac{f_{e\mu}}{M_{S_1}} \leq 10^{-4}{\rm GeV}^{-2}
\end{equation}
The Lagrangian $(6)$ can also lead to the flavor violating process
$\mu \rightarrow e + \gamma$. Using the experimental bound on its
branching ratio we find:
\begin{equation}
\frac{f_{e\alpha}f^*_{\mu \alpha}}{M_{S_1}^2} \leq
2.8.10^{-9}{\rm GeV}^{-2}
\end{equation}
The above constraints can be satisfied for the $f_{e\alpha} \sim
0.1$ and $M_{S_1} \sim $ few TeV, and are also consistent with limits from neutrinoless
double $\beta$-decay experiments, and are thus comparable with the requirements needed
to generate neutrino masses and dark matter, as described above.

Although the right-handed neutrino is a singlet under the SM gauge group, it may be produced in our
model through the scatterings of SM particles and the subsequent decay of $S_2$.

If the center of mass energy of the next generation linear collider exceeds twice the mass of $S_2$ then
these particles would be produced in, for example, $e^+ e^-$ annihilation. In this case, the dominant decay channel for the $S_2$ particle would be into a charged lepton and the lightest right-handed neutrino, providing a possible collider signature of the model.

\section{Conclusions}
It would be particularly exciting if a single mechanism might resolve two of the most important outstanding puzzles in particle physics and cosmology: the nature of dark matter, and the origin of neutrino masses.   As we demonstrate here, it is possible to extend standard model of particle physics in a way so that both puzzles may find a common resolution. The symmetry structure of our model is such that neutrino masses occur only at three-loop level, and so are naturally small. As a consequence, the right-handed neutrino is not constrained to be have a large mass, as in other examples of neutrino mass generation, such as the seesaw model. Such a light right-handed neutrino is a possible weakly interacting massive particle (WIMP) dark matter candidate.

We have explored the experimental constraints on our model and it currently passes all standard model tests.  More exciting perhaps is the fact that the TeV-scale scale energies required should be accessible at colliders in the near future. The model may also be tested by future neutrinoless
double $\beta$-decay experiments.  A new sort of direct detection WIMP technology would be required to directly detect the dark matter candidate we propose. 

\acknowledgements
LMK and SN are supported by the US Department of Energy (DOE)
LMK acknowledges the hospitality of the Kavli Institute for Theoretical Physics during this work. The work of MT is supported by the National Science
Foundation (NSF) under grant PHY-0094122.

\end{document}